\documentclass[twocolumn,twoside,slac_two]{revtex4}
\usepackage{graphicx}
\usepackage{fancyhdr}
\begin{document}

\title{The ATLAS Detector: Status and Results from Cosmic Rays}

\author{James T. Shank}
\affiliation{Physics Department , Boston University, Boston MA 02215, USA}

\author{On Behalf of the ATLAS Collaboration}

\begin{abstract}
The ATLAS detector at the Large Hadron Collider, CERN has been under construction for more than a decade. It is now largely complete and functional. This paper will describe the state of the major subsystems of ATLAS. Results from the brief single beam running period in 2008 will be shown. In addition, results from a long period of cosmic ray running will be shown. These results show that ATLAS is prepared to make major new physics discoveries as soon as we get colliding beams in late 2009.
\end{abstract}

\maketitle

\thispagestyle{fancy}


\section{Introduction}
The ATLAS detector at CERN is one of two large general purpose detectors at the Large Hadron Collider (LHC) \cite{LHC}, a 7 TeV on 7 TeV proton accelerator. It was designed to have excellent tracking, calorimetry and muon spectroscopy over the entire energy range of the LHC allowing discovery of any new physics in that range. Particular discovery potentials include details of the Standard Model and beyond, such as discovery of the Higgs boson, Supersymmetry (SUSY) and extra dimensions. More complete discussion of the physics discovery potential of ATLAS is given in reference \cite{atlas_performance}.

The ATLAS detector, under construction for well over a decade, was ready to record collision events on September 10, 2008 when the LHC started up. Single beams were circulated around the accelerator and resulting events were recorded in the detector. As is well known, about one week later, an incident in a superconducting splice between two dipoles resulted in a shutdown of the machine before any beam collisions occurred. This shutdown is still continuing and ATLAS has used this time to fix minor problems with the detector and commission the subsystems with cosmic ray events. The LHC accelerator is now scheduled to start again in November, 2009. This paper will describe the current state of each subsystem and present results of these cosmic ray events showing that ATLAS now fully ready for collision data that is expected later this year.

\section{Overview of the ATLAS Detector}

\begin{figure*}[t]
\centering
\includegraphics[width=135mm]{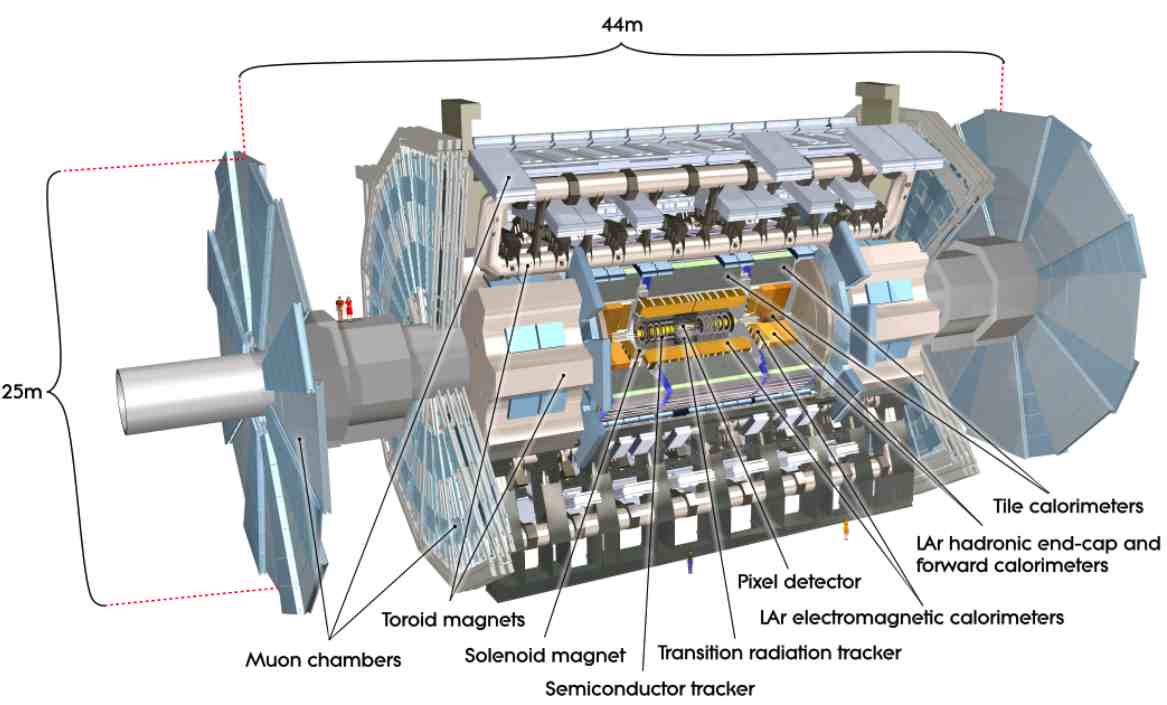}
\caption{A schematic view of the ATLAS detector.} \label{atlas_detector}
\end{figure*}

The ATLAS detector is large in more than one way: the outer dimensions (a cylinder 44m long and 25m in diameter) and the collaboration: over 2500 scientists from nearly 200 institutes in 37 countries.
An overview schematic of the detector is shown in~Fig.~\ref{atlas_detector} where the major subsystems are labelled.

%

\begin{figure}[h]
\centering
\includegraphics[width=80mm]{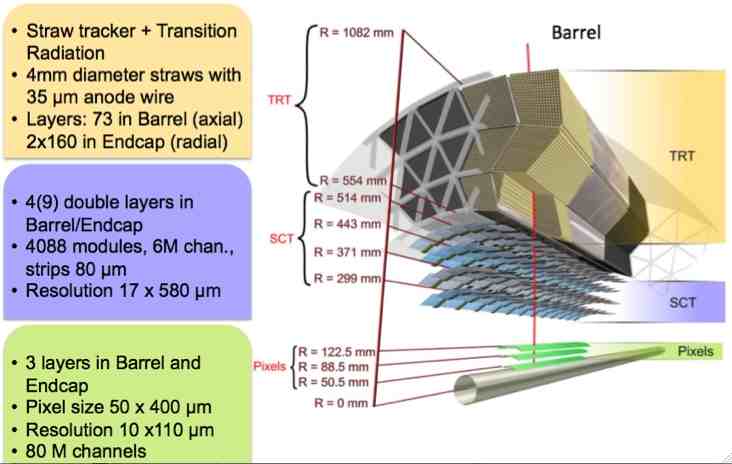}
\caption{A schematic view of the ATLAS inner detector.} \label{inner_detector}
\end{figure}

The outermost part of ATLAS is the muon spectrometer and its associated superconducting toroidal magnets which give ATLAS its name: A Toroidal LHC Apparatus. The key design in the muon spectrometer is minimal  material provided by having air-core toroidal magnets.  Next, moving inward is the calorimeter system consisting of three different technologies. The hadronic calorimeter is a scintillating tile/iron structure covering $\eta < 1.7$. In the endcap region $1.5 < \eta < 3.2$ hadronic calorimetry is provided by a Copper/Liquid Argon structure providing four longitudinal samples. Forward hadronic calorimetry is provided by a Tungsten/Liquid Argon structure covering $ 3.2 < \eta < 4.9$ with three longitudinal samples. Electromagnetic calorimetry is provided by a Lead/Liquid Argon detector  $ \eta < 2.5$ with three longitudinal samples together with a presampler detector in the  $ \eta < 1.8$ region. Inside of the calorimeter is the inner detector consisting of a transition radiation tracker (TRT), semiconductor tracker,  and a pixel detector shown in Fig.~\ref{inner_detector}.  The magnetic field for the inner detector is provided by a solenoid magnet providing a 2 Tesla field.  The TRT provides $ e-\pi $ separation over the energy range $ 0.5 < E < 150$ GeV. The pixel readout has 80 million channels, another one of the superlatives in ATLAS.

Complete details of the ATLAS detector are given in reference \cite{atlas_detector}

\section{Single Beam Events}

\begin{figure}[h]
\centering
\includegraphics[width=80mm]{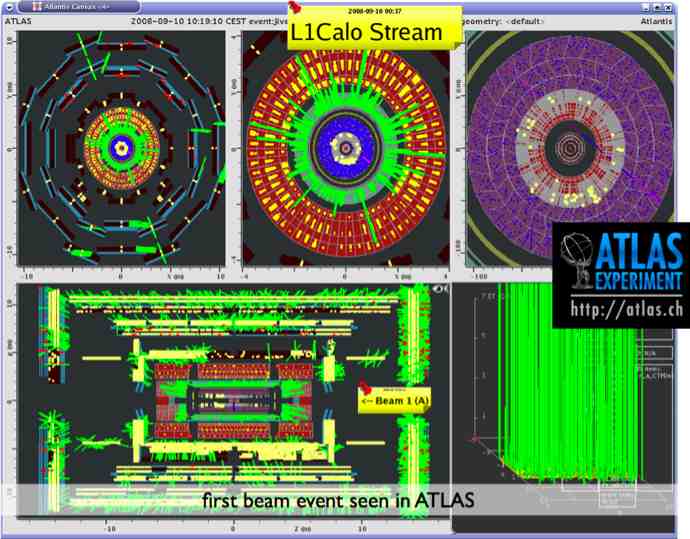}
\caption{ATLAS event display for a "splash" event.} \label{splash_event}
\end{figure}

During the brief single beam running in September, 2008, ATLAS recorded many events from beam halo (interactions with gas in the beam pipe) and so called beam-splash events produced when the beam was run into a collimator approximately 150m upstream of the detector. The beam splash events provided rather spectacular event display pictures, such as shown in Fig.~\ref{splash_event}, but also allowed time synchronization between the subsystems and the LHC machine.

\section{Cosmic ray running in 2008-9}

\begin{figure}[h]
\centering
\includegraphics[width=80mm]{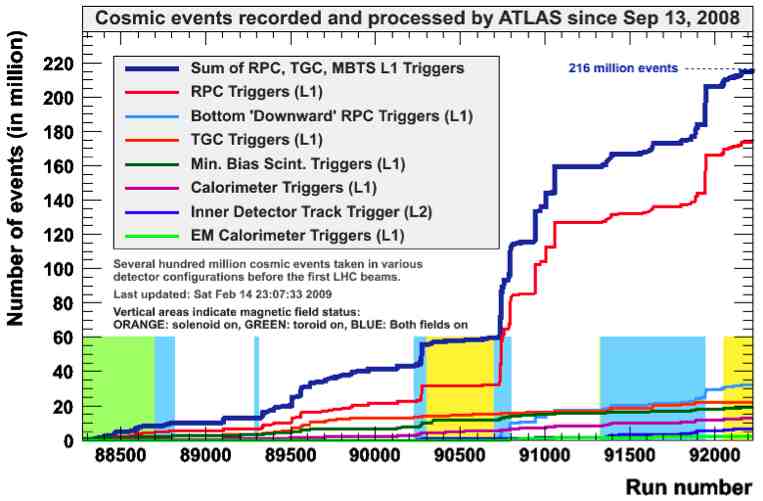}
\caption{Cosmic ray running in 2008, since the initial turn-on.} \label{cosmic_runs}
\end{figure}

Since the end of the beam in Sept. 2008, ATLAS has been accumulating cosmic ray events in all subsystems of the detector. Fig.~\ref{cosmic_runs} shows the running periods and the state of the magnets during each run. Over 200 million events have been recorded.
%

The following sections will review the status of each ATLAS subsystem and show results from these cosmic ray runs.
\medskip

\section{The Inner Detector}

\bigskip

\begin{figure}[h]
\centering
\includegraphics[width=80mm]{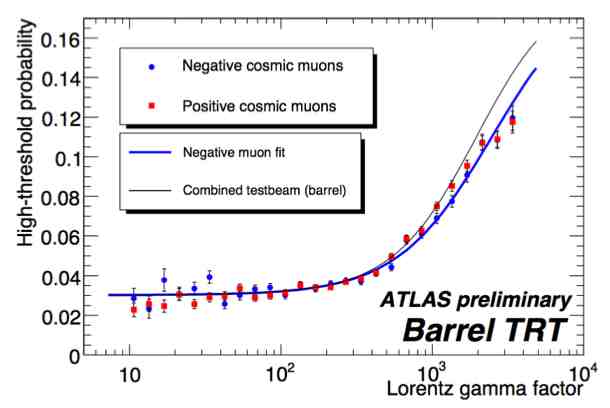}
\caption{Muon identification in the Transition Radiation Tracker.} \label{trt_muons}
\end{figure}

The ATLAS Inner Detector has three major detector components: the pixel detector, the silicon detector and the transition radiation tracker as shown above in Fig.~\ref{inner_detector}.
The pixel detector currently has approximately $98 \%$ live channels, a hit efficiency of $\approx 99.8\% $ and a noise occupancy of $\approx 10^{-10} $ The Semiconductor Tracker has over $ 99\%$ of its 6 million channels operational. The noise occupancy is $1.5 \times 10^{-5}$ in the barrel region and $ 3.0 \times 10^{-5}$ in the endcap. The Transition Radiation Tracker (TRT) has $98\%$ of its 350k channels operational.  The transition radiation aspects of the TRT have been proven in testbeam studies where the full turn on of transition radiation is seen and in the cosmic ray runs where the beginning of the turn on is seen. This is shown in Fig.~\ref{trt_muons} where the probability of a signal above the high threshold is shown as a function of the Lorentz gamma factor of the cosmic muon. Transition radiation photons produced by particles with high gamma-factor cause a larger energy deposition in the straw tubes of the \penalty 1000 TRT.

\section{The Calorimeter System}

\begin{figure}[h]
\centering
\includegraphics[width=80mm]{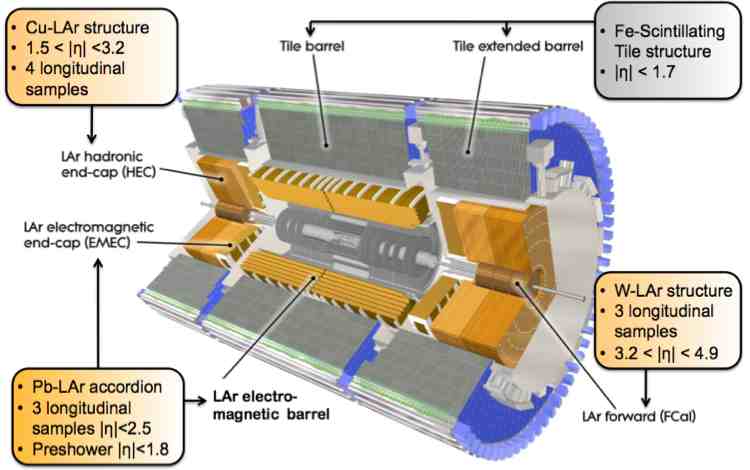}
\caption{The ATLAS calorimeter system.} \label{calorimeter}
\end{figure}

The ATLAS Liquid Argon (LAr) calorimeter systems, described above and shown in more detail in Fig.~\ref{calorimeter}, has $98.8 \% $ of channels operational.  The Hadronic Tile Calorimeter has $0.4\%$ dead channels, which will be fixed at the next shutdown. The electronic calibration systems for all the calorimeter systems are fully operational.

\begin{figure}[h]
\centering
\includegraphics[width=80mm]{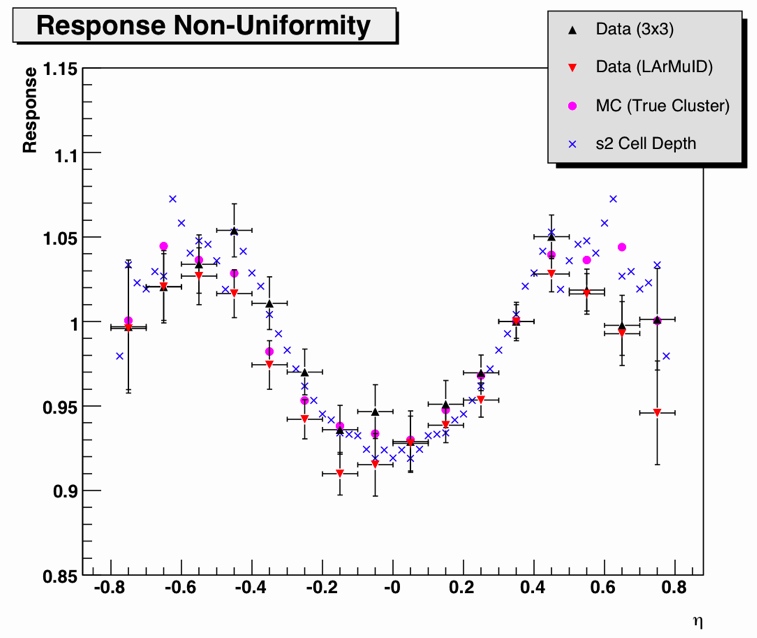}
\caption{Signal response to cosmic rays in the Liquid Argon Electromagnetic Calorimeter. The plot shows the most probable energy normalized to the $\eta$ point between 0.3 and 0.4} \label{lar_response}
\end{figure}

An example of studies with cosmic rays is shown in Fig.~\ref{lar_response}, where the response of the LAr barrel electromagnetic calorimeter is shown as a function of pseudorapidity. Here the response is the most probable value (MPV) determined by two different algorithms. The LArMuID algorithm is a variable size algorithm - only cells above a given threshold are added to the cluster.  The 3x3 cluster algorithm  is fixed in size.  Also shown are the true cluster response from Monte Carlo simulation and the cell depth. The data match the simulation and follow the cell depth as expected. The uniformity of response agrees with the simulation at the $2\% $ level.

\section{The Muon Spectrometer}

\begin{figure}[h]
\centering
\includegraphics[width=80mm]{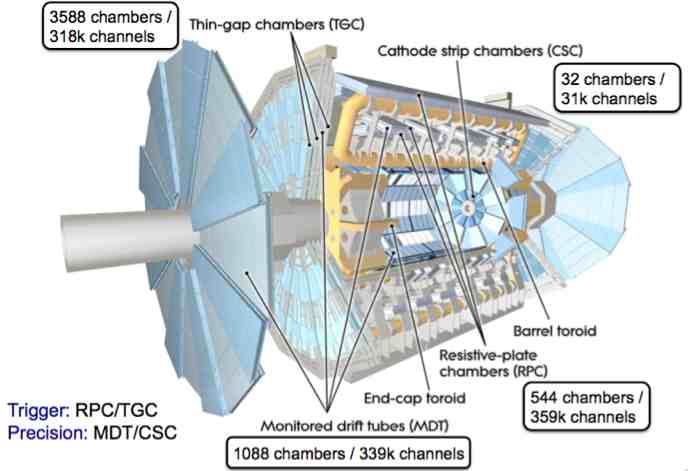}
\caption{The ATLAS Muon Spectrometer. } \label{muon_detector}
\end{figure}

The Muon Spectrometer in ATLAS has two different technologies for the precision measurement and two for the trigger as shown in Fig.~\ref{muon_detector}. The Monitored Drift Tube (MDT) system together with the Cathode Strip Chambers (CSC) in the forward region provide a spatial resolution of $35-40~\mu m$. The MDT tubes are "monitored" by a 12232 channel optical alignment system which provides position accuracy of $30~\mu m$. The MDT system has $0.3\%$ dead channels. The CSC system has $1.5\%$ dead channels.

The muon trigger is provided by Resistive Plate Chambers (RPC) in the barrel and Thin Gap Chambers (TGC) in the endcap. These provide a spatial resolution of $5-10$mm and a time resolution sufficient to allow bunch crossing identification with the $25$ ns bunch crossing time of the LHC. The RPC system has $95.5\%$ channels operational with an additional $3 \%$ recoverable during a shutdown. The TGC system is  $99.8\%$ operational with less than $0.02\%$ noisy channels. 

Overall, the muon system provides a stand alone (that is, not combined with inner detector measurements) resolution of $\Delta P_T/P_T < 10\%$ up to 1 TeV.

\begin{figure}[h]
\centering
\includegraphics[width=80mm]{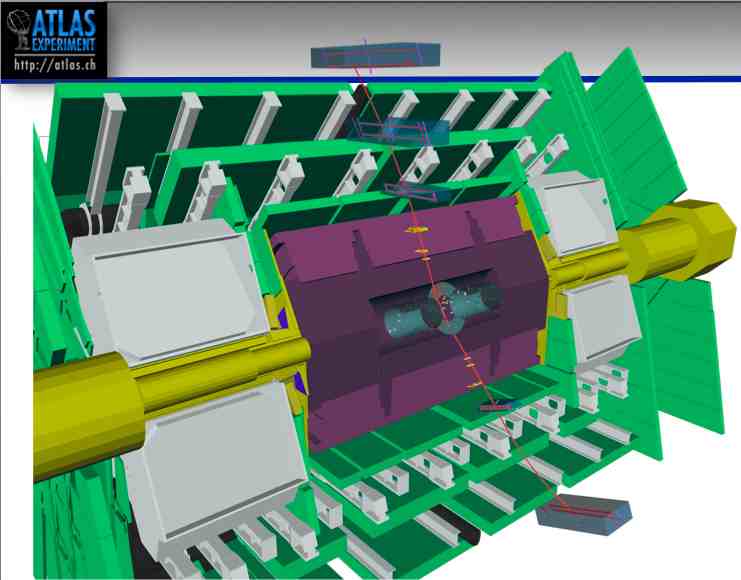}
\caption{The ATLAS event display showing a cosmic ray muon passing through the entire detector. Only the muon chambers with hits are shown for visibility. } \label{muon_event}
\end{figure}

A typical cosmic ray muon seen in the spectrometer is shown in Fig.~\ref{muon_event}. As can been seen by the track bending, the magnetic field was on for this event.

\begin{figure}[h]
\centering
\includegraphics[width=80mm]{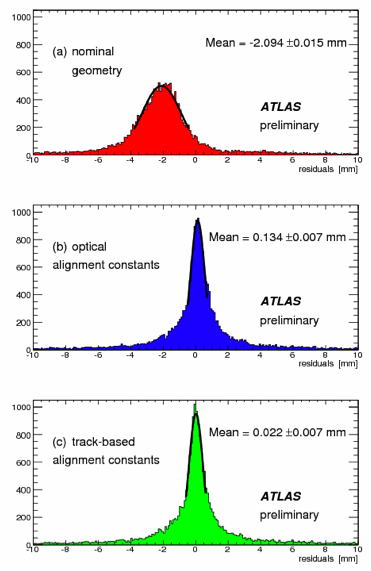}
\caption{Track residuals in the Muon Spectrometer showing the effect of alignment. a) nominal geometry with no alignment. b) corrections from the optical alignment system applied. c) constraints from track-based alignment applied. } \label{muon_residuals}
\end{figure}

Effects of the muon alignment system can be seen in Fig.~\ref{muon_residuals} where the sagitta measured in the bend plane of the magnet is shown from tracks that go through one particular chamber in the muon barrel system. The top plot in that figure shows the sagitta with just the nominal geometry for the chambers that the track passes through. The middle plot shows the sagitta after corrections given by the optical alignment system. The bottom plot shows the final distribution obtained after applying corrections from track-based alignment. 

\section {Electrons in Cosmic Ray Data}

For this analysis, 3.5 million cosmic ray events which passed the high-level trigger track reconstruction in the barrel inner detector were used.  After filtering for electromagnetic cluster candidates with transverse energy above $\sim$3 GeV and with a loose track match in $\phi $   (pointing downwards), about 11 000 candidates remain. 
These remaining events are required to satisfy medium electron cuts (lateral shower 
shapes in first and second layers of EM calorimeter and track- 
cluster matching in $\phi$ for tracks with at least 25 TRT hits) and are 
then split into two categories: 
a) a sample consisting of 1229 muon bremsstrahlung candidates, 
with only one track reconstructed in the barrel inner detector 
b) a sample consisting of 85 ionisation electron candidates, with at 
least two tracks reconstructed in the barrel inner detector

\begin{figure}[h]
\centering
\includegraphics[width=80mm]{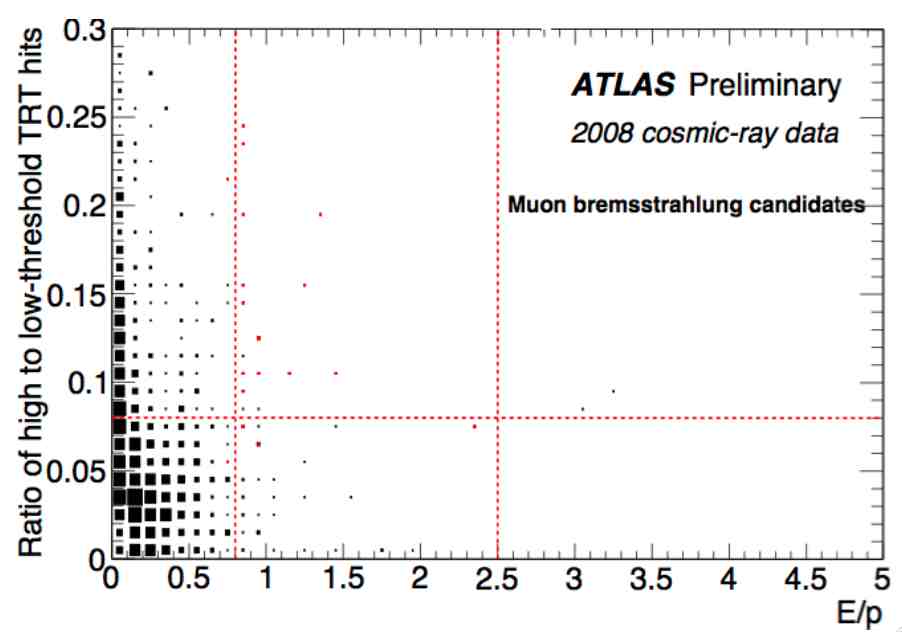}
\caption{Ratio of high to low-threshold TRT hits versus E/p, where E is measured in 
the EM calorimeter and p in the inner detector, for bremsstrahlung candidates.} \label{electrons_1}
\end{figure}

\begin{figure}[h]
\centering
\includegraphics[width=80mm]{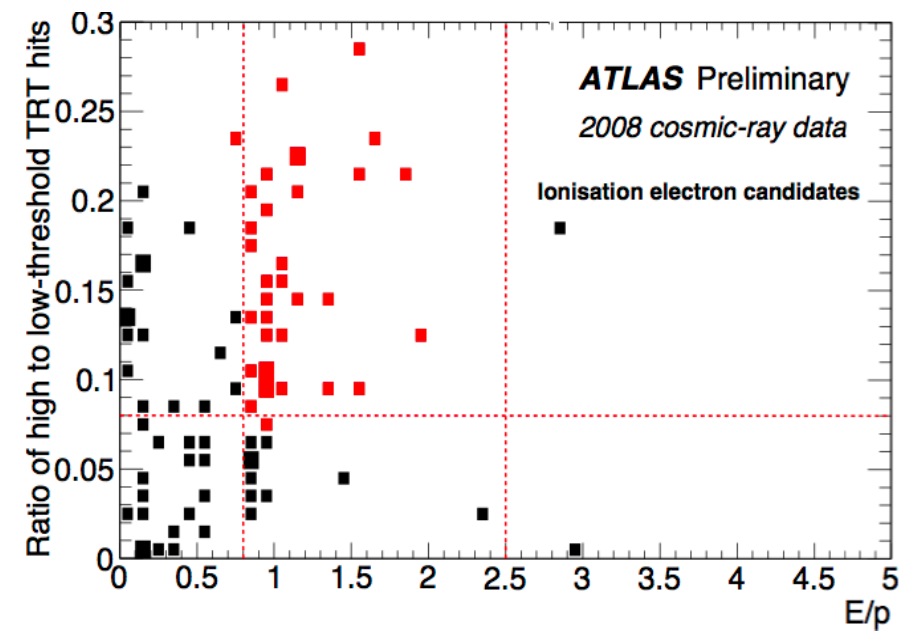}
\caption{Ratio of high to low-threshold TRT hits versus E/p, where E is measured in 
the EM calorimeter and p in the inner detector, for ionization electron candidates.} \label{electrons_2}
\end{figure}

In Fig.~\ref{electrons_1}, the red boxes correspond to candidates satisfying additional cuts defined for standard tight high$-p_T$ electron identification in ATLAS at $\eta\sim 0$. These cuts 
are $p_T$ and $\eta$-dependent and the ones applied to most of the events are 
illustrated by the dashed red lines in figures~\ref{electrons_1} and \ref{electrons_2}:
$ 0.8 < E/p < 2.5$ and high to 
low-threshold TRT hit ratio $> 0.08$ (indicating the detection of transition radiation produced only by relativistic particles). 
Most of the events in the muon bremsstrahlung sample have small E/p and few 
high-threshold TRT hits (only muons above 100 GeV momentum are expected 
to produce transition radiation). The events with low E/p and high TRT ratio 
contain a large fraction of muons with combined momentum measured in the 
muon spectrometer and ID above 100 GeV. Only 19 of the 1229 events satisfy 
the signal criteria. 
In contrast, in the electron ionisation sample, a large fraction of events, 36 out 
of a total of 85, satisfy the signal criteria. These events are interpreted as high- 
energy $\delta$-rays produced in the inner detector volume by the incoming cosmic 
muons.

\begin{figure}[h]
\centering
\includegraphics[width=80mm]{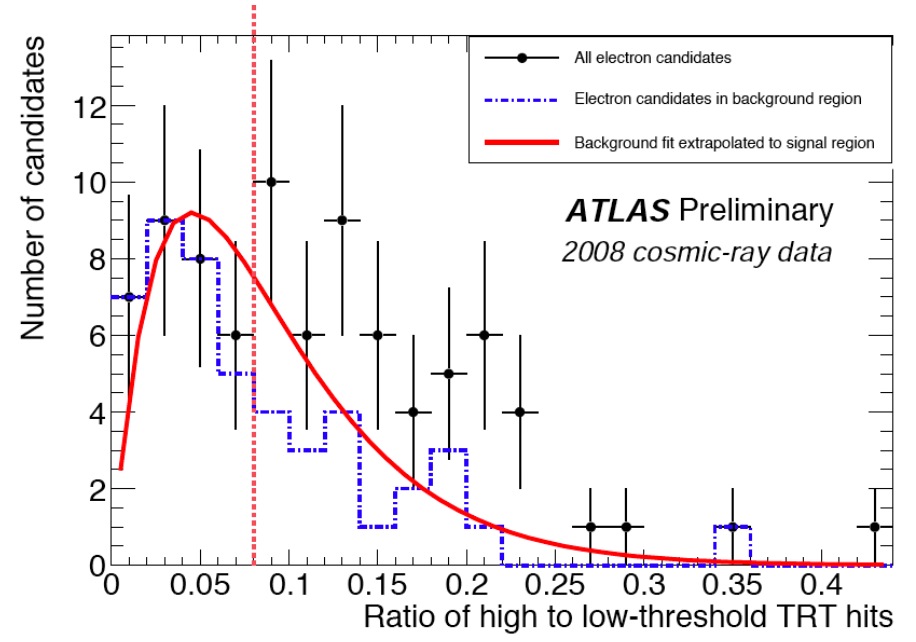}
\caption{Distribution of the ratio of the high to low-threshold TRT hits for the 
85 ionisation electron candidates (black) and for the 49 candidates failing 
the final signal criteria (blue).} \label{electrons_3}
\end{figure}

In Fig.~\ref{electrons_3}, the red curve is the projection of a two dimensional binned maximum likelihood fit to the two dimensional plot in Fig.~\ref{electrons_2}, excluding the signal candidates. The shape of the projected distributions is obtained from the muon bremsstrahlung sample, but the parameters are fitted using the ionisation sample. This fit is used 
as one of the methods to estimate the background contamination to the 
signal. 
A clear excess of events with a large ratio of high to low-threshold TRT 
hits is observed, indicating the presence of an electron signal.

\begin{figure}[h]
\centering
\includegraphics[width=80mm]{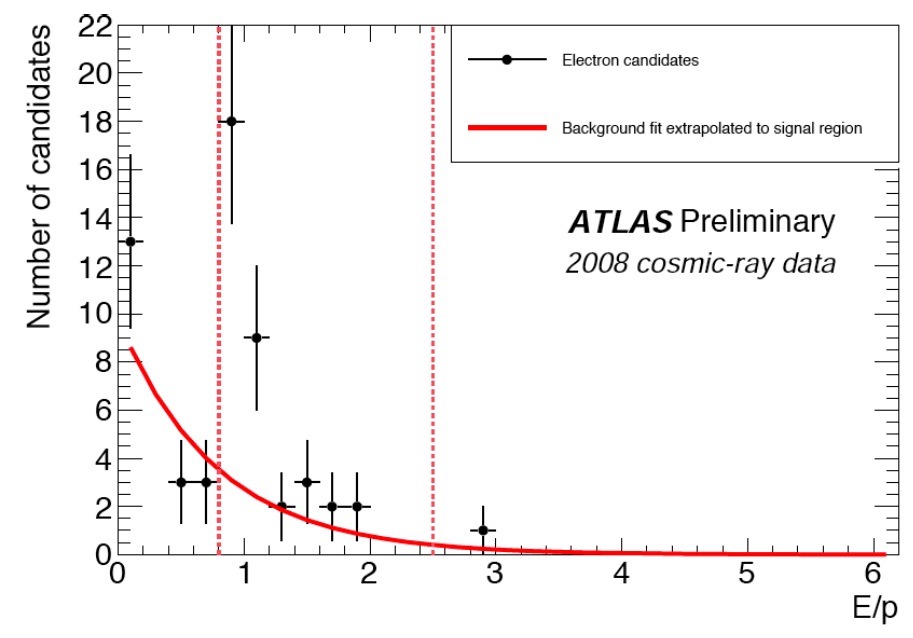}
\caption{Distribution of the energy to momentum ratio, E/p, for the electron candidates of 
Fig.~\ref{electrons_2}, after applying the cut indicated on the ratio of high to low-threshold TRT 
hits. The red curve shows the background, estimated as explained in Fig.~\ref{electrons_3} and 
projected on the E/p distribution. The background under the signal is estimated to 
be ($8.7 \pm  3.1$) events, using this first method described above for Fig.~\ref{electrons_3}.} \label{electrons_4}
\end{figure}

Figure~\ref{electrons_4}, shows the Distribution of energy to momentum ratio for the electrons candidates. A clear accumulation of signal events around E/p = 1 is observed, as expected for 
electrons. 

A second independent method to estimate the background uses the measured ratio 
of negatively to positively charged muons coupled to the fact that the electron 
ionisation signal should contain no positrons: 
- the $\mu^-/\mu^+$  ratio obtained from the muon bremsstrahlung sample is $\sim 0.70$ 
- out of the 36 signal candidates, four have a positive charge, leading to an 
expectation of ($6.8 \pm 3.4$) background events in the signal sample, in good 
agreement with the estimate from the first method. 
The final sample consists of the 32 candidates with measured negative charge. 

\section{Conclusions}
Due to space limitations, some systems of ATLAS have been omitted from  this paper. Notably the trigger system which has also been commissioned during this period and is also ready for fist collisions. In addition, the ATLAS distributed computing system, with its world-wide computing infrastructure, a crucial component of the analysis and data processing system, has been undergoing continuous commissioning. 

In summary, all the subsystems of ATLAS have demonstrated capability sufficient to ensure physics discoveries as soon as collision data is ready later this year.

\begin{acknowledgements}

We are greatly indebted to all CERN's departments and to the LHC project for their immense efforts not only in building the LHC, but also for their direct contributions to the construction and installation of the ATLAS detector and its infrastructure. We acknowledge equally warmly all our technical colleagues in the collaborating Institutions without whom the ATLAS detector could not have been built. Furthermore we are grateful to all the funding agencies which supported generously the construction and the commissioning of the ATLAS detector and also provided the computing infrastructure.

The ATLAS detector design and construction has taken about fifteen years, and our thoughts are with all our colleagues who sadly could not see its final realisation.

We acknowledge the support of ANPCyT, Argentina; Yerevan Physics Institute, Armenia; ARC and DEST, Australia; Bundesministerium f\"ur Wissenschaft und Forschung, Austria; National Academy of Sciences of Azerbaijan; State Committee on Science \& Technologies of the Republic of Belarus; CNPq and FINEP, Brazil; NSERC, NRC, and CFI, Canada; CERN; NSFC, China; Ministry of Education, Youth and Sports of the Czech Republic, Ministry of Industry and Trade of the Czech Republic, and Committee for Collaboration of the Czech Republic with CERN; Danish Natural Science Research Council; European Commission, through the ARTEMIS Research Training Network; IN2P3-CNRS and Dapnia-CEA, France; Georgian Academy of Sciences; BMBF, DESY, DFG and MPG, Germany; Ministry of Education and Religion, through the EPEAEK program PYTHAGORAS II and GSRT, Greece; ISF, MINERVA, GIF, DIP, and Benoziyo Center, Israel; INFN, Italy; MEXT, Japan; CNRST, Morocco; FOM and NWO, Netherlands; The Research Council of Norway; Ministry of Science and Higher Education, Poland; GRICES and FCT, Portugal; Ministry of Education and Research, Romania; Ministry of Education and Science of the Russian Federation, Russian Federal Agency of Science and Innovations, and Russian Federal Agency of Atomic Energy; JINR; Ministry of Science, Serbia; Department of International Science and Technology Cooperation, Ministry of Education of the Slovak Republic; Slovenian Research Agency, Ministry of Higher Education, Science and Technology, Slovenia; Ministerio de Educaci\'{o}n y Ciencia, Spain; The Swedish Research Council, The Knut and Alice Wallenberg Foundation, Sweden; State Secretariat for Education and Science, Swiss National Science Foundation, and Cantons of Bern and Geneva, Switzerland; National Science Council, Taiwan; TAEK, Turkey; The Science and Technology Facilities Council and The Leverhulme Trust, United Kingdom; DOE and NSF, United States of America. 

\end{acknowledgements}


\end{document}